\documentclass[prl,twocolumn,superscriptaddress]{revtex4}

\usepackage{pslatex}
\usepackage{amsmath}
\usepackage{amssymb}
\usepackage{graphicx}
\usepackage{pst-node}
\usepackage{pst-coil}

\begin{document}
\bibliographystyle{apsrev}

\title{Detecting Non-Abelian Statistics in the
$\nu=5/2$ Fractional Quantum Hall State}


\author{Parsa Bonderson}
\email[]{pbonders@theory.caltech.edu}
\affiliation{California Institute of Technology, Pasadena, CA 91125}
\author{Alexei Kitaev}
\email[]{kitaev@iqi.caltech.edu}
\affiliation{California Institute of Technology, Pasadena, CA 91125}
\author{Kirill Shtengel}
\email[]{kirill@caltech.edu}
\affiliation{California Institute of Technology, Pasadena, CA 91125}
\affiliation{Department of Physics, University of California,
Riverside, CA 92521}

\date{\today}

\begin{abstract}
  In this Letter we propose an interferometric experiment to detect
  non-Abelian quasiparticle statistics -- one of the hallmark
  characteristics of the Moore-Read  state
  expected to describe the observed FQHE plateau at $\nu= 5/2$. The
  implications for using this state for constructing a topologically
  protected qubit as has been recently proposed by Das Sarma \emph{et.
    al.} are also addressed.
\end{abstract}

\maketitle

\paragraph{Introduction}
One of the most interesting aspects of the Fractional Quantum Hall
Effect (FQHE) is the fractionalized nature of its quasiparticle excitations.
In addition to carrying a fraction of the electron charge, these
excitations are generally expected to have exotic exchange statistics
which are neither bosonic nor fermionic. These exotic statistics,
generally allowed in (2+1)-dimensions \cite{Leinaas77}, are given by
representations of the braid group (as opposed to higher dimensions
where statistics is represented by the permutation group), and
particles that transform as such have been dubbed anyons
\cite{Wilczek82a,Wilczek82b}.
The fractional charge of quasiparticles in the $\nu =1/3$ Laughlin
state was first measured a decade ago \cite{Goldman95}, but
confirmation of their statistics remained elusive until very recently
\cite{Goldman05a,Camino05a,Camino05b}. Aside from the experimental
difficulties associated with measuring quasiparticle interference
patterns, there are also conceptual issues regarding how to isolate
the contribution of braiding statistics from that of the Aharonov-Bohm
phases that arise due to the quasiparticle charge encircling a region
of magnetic flux. For a careful discussion of this subject, see
\cite{Chamon97}. Curiously, isolating these pieces may prove easier in
a more exotic state with non-Abelian statistics. In such a system, the
Hilbert space is multi-dimensional and exchange transformations may
rotate different states into one another.  This notion, along with a
topological protection inherent in these systems make them attractive
candidates for fault-tolerant quantum computation
\cite{Kitaev97,Freedman02a,Preskill-lectures}. A concrete proposal for
creating a topologically protected qubit has been recently put forward
in \cite{DasSarma05}.

While the existence of Abelian anyons has been well established in the
context of FQHE, the more exciting prospect that non-Abelian anyons
exist has not been experimentally confirmed. The prime candidate for
finding non-Abelian statistics seems to be the FQH state observed at
the $\nu =5/2$ plateau \cite{Willett87,Pan99,Eisenstein02,Xia04}.
While its first Landau level counterpart, the $\nu =1/2$ state, is
widely believed to be a Fermi liquid of composite fermions
\cite{Halperin93}, it is most likely that the $\nu =5/2$ system is the
$p$-wave (spin-polarized) superconducting condensate described by the
Moore-Read (MR), \emph{aka} Pfaffian, state \cite{Moore91,Greiter92}.
Experimental evidence of spin-polarization \cite{Pan01}, together with
careful numerical studies \cite{Morf98,Rezayi00}, indicate a preference
for the MR state over other potential candidates, notably the
Abelian (3,3,1) Halperin state \cite{Halperin83}, the non-Abelian
(albeit critical) Haldane-Rezayi state \cite{Haldane88} and the
compressible striped phase \cite{Koulakov96}.

\paragraph{Proposed experimental setup}

\begin{figure}[hbt]
\includegraphics[width=1.8in]{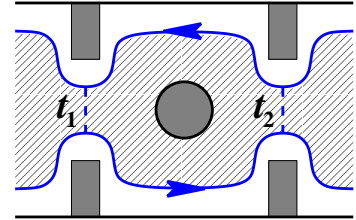}
\caption{A two point-contact interferometer for measuring the quasiparticle
statistics. The hatched region contains an
  incompressible FQH liquid. The front gates (grey rectangles) are
  used to bring the opposite edge currents (indicated by arrows) close
  to each other to form two tunneling junctions.  Applying voltage to
  the central gate creates an antidot in the middle and controls the
  number of quasiparticles contained there.}
\label{fig:interferometer}
\end{figure}

The experimental device we would like to consider is a two
point-contact interferometer composed of a quantum Hall bar with two
front gates on either side of an antidot (see
Fig.~\ref{fig:interferometer}). 
Biasing the front gates can be used to create constrictions in the
Hall bar, adjusting the tunneling amplitudes $t_{1}$ and $t_{2}$. The
relative amplitudes can be compared by individually switching them on.
The tunneling between the opposite edge currents leads to the
deviation of $\sigma_{xy}$ from its quantized value, or equivalently,
to the appearance of $\sigma_{xx}$.  The goal of the experiment is to
observe the interference between the two tunneling paths that the
quasihole current may traverse.  For this experiment, we are
interested in the \emph{weak} backscattering regime, i.e. the case
where the tunneling amplitudes $t_{1}$ and $t_{2}$ are small. The main
reason for this is to ensure that the tunneling current is entirely
due to charge $e/4$ quasiholes (with essentially no contribution from
the higher charge composites), which is a crucial component of our
predictions. In this regime, such tunneling is indeed the most
relevant perturbation \cite{Kane94,Wen92b}, but this need not be true
in the strong tunneling regime, where the constrictions are
effectively pinched off.  We should also mention that interpreting the
interference pattern is simplified when $t_1$ and $t_2$ are small.

For the purposes of this experiment, we envision three main
experimentally variable parameters: (i) the central gate voltage
allowing one to control the number $n$ of quasiholes on the antidot,
(ii) the magnetic field $B$, and (iii) the back gate voltage
controlling the uniform electron density.  This setup is essentially
identical to that proposed for measuring statistics in the Abelian
FQHE \cite{Chamon97}, later adopted for the non-Abelian case in
\cite{Fradkin98}, and not dissimilar to the one experimentally
realized in \cite{Goldman05a,Goldman05bd}.

To lowest order in $t_{1}$ and $t_{2}$, the tunneling current and, hence,
longitudinal conductivity $\sigma _{xx}$ in this system will be
proportional to the probability that current entering the bottom edge
leaves through the top edge:
\begin{eqnarray}
\sigma _{xx} 
&\varpropto &\left| \left( t_{1}U_{1}+t_{2}U_{2}\right) \left| \Psi
\right\rangle \right| ^{2}  \notag \\
&=&\left| t_{1}\right| ^{2}+\left| t_{2}\right| ^{2}+2\text{Re}\left\{
t_{1}^{\ast }t_{2}\left\langle \Psi \right| U_{1}^{-1} U_{2}\left| \Psi
\right\rangle \right\}   \notag \\
&=&\left| t_{1}\right| ^{2}+\left| t_{2}\right| ^{2}+2\text{Re}\left\{
t_{1}^{\ast }t_{2}e^{i\alpha }\left\langle \Psi \right| M_{n}\left| \Psi
\right\rangle \right\} 
\end{eqnarray}
In this expression, $U_{1}$ and $U_{2}$ are the unitary evolution
operators for a quasihole taking the two respective paths, and $\left|
  \Psi \right\rangle $ is the initial state of the system. In the
third line, $e^{i\alpha }$ is the phase acquired from the dynamics of
traveling along the edge around the center region together with the
Aharonov-Bohm phase from taking the quasihole charge around the
magnetic flux through the center region. The operator $M_{n}$ is the
transformation due solely to the braiding statistics of winding a
single quasiparticle around $n$ quasiparticles. Its value for the MR
state was related to the Jones polynomial in \cite{Fradkin98} using
the Chern-Simons effective theory. We shall extend their analysis and
show explicitly how to detect the non-Abelian statistics. 

If we keep the filling factor fixed by simultaneously adjusting $B$
and the electron concentration, so as to keep the the quasihole number
constant, the Aharonov-Bohm phase as a function of
$\Phi $ will have a periodicity of $({e}/{e^{\ast}})\Phi _{0}$ where $\Phi
_{0}=2\pi/e$ (in units $\hbar=c=1$) and $e^{\ast }$ is the
electric charge of the quasiholes \cite{Chamon97}. Thus, varying the
flux $\Phi $ allows one to determine the quasiparticle charge. Note
that for $\nu=5/2$, a quasihole charge of $e^{\ast }= e/4$ rather than
$e^{\ast }=e/2$, would be indicative of a paired state.

\paragraph{The Moore-Read state}

The braiding statistics of particles in a (2+1)-dimensional quantum
system may be described by a general model of anyons (see
\cite{Preskill-lectures,Kitaev05} and references therein). Such a
model is defined by a set of particle types, fusion rules, and
braiding rules, all of which are required to satisfy certain
consistency conditions. The particle types and their fusion rules can
be respectively thought of as generalizations of group representations
and their tensor products, specifying values of conserved charges and
the possible values that may be obtained when forming composite
objects (composite in this context need not necessarily mean that the
constituents are bound together, but simply that their local
properties are not being individually probed, as in the case when they
are being viewed from far away).

The anyon model that describes the MR state can be denoted as
$\text{U}\left( 1\right) \times \text{Ising}$. (The term `Ising' is
used here in reference to the anyon model obtained from the
holomorphic part of the conformal field theory that describes the
Ising model at criticality.) In this notation, $\text{U}\left(
  1\right) $ refers to the familiar Abelian charge/flux sector, for
which particle type is specified by the amount of charge and flux
carried, the fusion rules are simply addition of these quantities
(i.e. the conservation of charge and flux), and the braiding rules are
specified by the usual phases acquired from winding charge/flux
composites, i.e. winding one $\left(q,\phi \right)$ charge/flux
composite around another produces a phase of $e^{iq\phi}$
\footnote{There is a subtle distinction between the statistics of
  $\text{U}(1)$ Chern-Simons theory, which is the proper description
  for FQH systems, and the usual $\text{U}(1)$ QED. In Chern-Simons
  theory, assigning a charge to a particle induces the flux, and the
  resulting statistical phase is half what one would expect from the
  electromagnetic Aharonov-Bohm effect.}.  Though less familiar, the
Ising anyon model (which contributes all of the non-Abelian statistics
to the MR state), is fairly simple.  It has three particle types,
conventionally denoted as: $\mathbb{I}$ (vacuum), $\sigma$
(spin/vortex), and $\psi$ (Majorana fermion)\footnote{To avoid
  confusion, we point out that there is an alternative way of treating
  these excitations \cite{Read00,Ivanov01,Stern04} - namely, by
  identifying $\sigma$ quasiparticles with the zero-energy Majorana
  modes inside a vortex core in a $p$-wave superconductor. In this
  language, $\psi$ particles (Majorana fermions in our notation)
  become usual fermionic modes. While we make no recourse to this
  description, it is worth mentioning that it provides an alternative
  derivation of the non-Abelian braiding rules, up to a phase
  factor.}, and the following fusion rules:
\begin{equation}
  \label{eq:fusion1}
  \begin{array}{ccc}
    \mathbb{I}\times \mathbb{I}=\mathbb{I}, & 
    \mathbb{I} \times \sigma =\sigma,  & 
    \mathbb{I} \times \psi =\psi,  \\ 
    \sigma \times \sigma = \mathbb{I} +\psi, \quad  & 
    \sigma \times \psi =\sigma, \quad  & 
    \psi \times \psi = \mathbb{I}.
  \end{array}
\end{equation}
In words, combining the two particle types on the left hand side of
the equal sign gives some superposition of states carrying the labels
on the right hand side (as mentioned earlier, one may think of the
symbols $\times$ and $+$ respectively as generalizations of the
tensor product, $\otimes$, and the direct sum, $\oplus$).
Graphically, these rules can be represented by locally permitting only
the following set of trivalent vertices (in any desired orientation):
\begin{equation*}
  \label{eq:fusion2}
  \pspicture[0.0](0.866,0.866)
  \psline[linewidth=1.5pt,linecolor=magenta] (0.0,0.0)(0.5,0.433)
  \psline[linewidth=1.5pt,linecolor=magenta] (0.5,0.433)(1.0,0.0)
  \pscoil[coilarm=1.3pt,coilwidth=5pt,coilaspect=0,linewidth=1.5pt,linecolor=blue]
  (0.5,0.433)(0.5,1.0)
  \rput[bl]{0}(-0.1,0.25){$\sigma$}
  \rput[bl]{0}(0.88,0.25){$\sigma$}
  \rput[bl]{0}(0.7,0.7){$\psi$}
\endpspicture
  \;  \; \; , \qquad
  \pspicture[0.0](0.866,0.866)
  \psline[linewidth=1.5pt,linecolor=magenta] (0.0,0.0)(0.5,0.433)
  \psline[linewidth=1.5pt,linecolor=magenta] (0.5,0.433)(1.0,0.0)
\psline[linewidth=1.4pt,linecolor=black,linestyle=dotted]
  (0.5,0.433)(0.5,1.0)
  \rput[bl]{0}(-0.1,0.25){$\sigma$}
  \rput[bl]{0}(0.88,0.25){$\sigma$}
  \rput[bl]{0}(0.68,0.75){$\mathbb{I}$}
  \endpspicture
   \;  \; \; , \qquad
  \pspicture[0.0](0.866,0.866)
  \pscoil[coilarm=1.3pt,coilwidth=5pt,coilaspect=0,linewidth=1.5pt,linecolor=blue]
  (0.0,0.0)(0.5,0.433)
  \pscoil[coilarm=1.3pt,coilwidth=5pt,coilaspect=0,linewidth=1.5pt,linecolor=blue]
  (0.5,0.433)(1.0,0.0)
\psline[linewidth=1.4pt,linecolor=black,linestyle=dotted]
  (0.5,0.433)(0.5,1.0)
  \rput[bl]{0}(-0.1,0.3){$\psi$}
  \rput[bl]{0}(0.88,0.3){$\psi$}
  \rput[bl]{0}(0.68,0.75){$\mathbb{I}$}
  \endpspicture
   \;  \; \; , \qquad
  \pspicture[0.0](0.866,0.866)
  \psline[linewidth=1.4pt,linecolor=black,linestyle=dotted]
  (0.0,0.0)(0.5,0.433)
  \psline[linewidth=1.4pt,linecolor=black,linestyle=dotted]
  (0.5,0.433)(1.0,0.0)
\psline[linewidth=1.4pt,linecolor=black,linestyle=dotted]
  (0.5,0.433)(0.5,1.0)
  \rput[bl]{0}(-0.1,0.25){$\mathbb{I}$}
  \rput[bl]{0}(0.88,0.25){$\mathbb{I}$}
  \rput[bl]{0}(0.68,0.75){$\mathbb{I}$}
  \endpspicture
\end{equation*}
From the anyon model rules and consistency conditions (which we will
not present here, but instead refer the reader to \cite{Kitaev05} for
details \footnote{The properties of the Ising anyon model can be found
  in Table 1 of \cite{Kitaev05} by replacing $\varepsilon$ with $\psi$ and
setting $v=1$.}), we can distill the following braiding rules:
\begin{subequations}
\label{eq:braiding}
\begin{equation}
  \label{eq:braiding1}
  \pspicture[0.5](1.8,2.0)
  \psline[linewidth=1.4pt,linecolor=black,linestyle=dotted]
  (1,0.0)(1.0,2.0)
  \pscurve[linewidth=1.5pt,linecolor=magenta,border=4pt]
  (0.2,0.0)(1,0.5)(1.4,1.1)(1.2,1.4)
  \pscurve[linewidth=1.5pt,linecolor=magenta](0.8,1.65)(0.5,1.85)(0.2,2.0)
  \rput[bl]{0}(-0.2,0.0){$\sigma$}
  \rput[bl]{0}(1.2,0){$\mathbb{I}$}
  \endpspicture
  \; =  \;
  \pspicture[0.5](2,2.0)
  \psline[linewidth=1.4pt,linecolor=black,linestyle=dotted]
  (1.2,0.0)(1.2,2.0)
  \psline[linewidth=1.5pt,linecolor=magenta](0.4,0.0)(0.4,2.0)
  \rput[bl]{0}(0,0.0){$\sigma$}
  \rput[bl]{0}(1.4,0){$\mathbb{I}$}
  \endpspicture
\end{equation}
\begin{equation}
  \label{eq:braiding2}
  \pspicture[0.5](1.8,2.0)
 
\pscoil[coilarm=1pt,coilwidth=5pt,coilaspect=0,linewidth=1.5pt,linecolor=blue]
    (1,0.0)(1.0,2.0)
  \pscurve[linewidth=1.5pt,linecolor=magenta,border=4pt]
  (0.2,0.0)(1,0.5)(1.4,1.1)(1.2,1.4)
  \pscurve[linewidth=1.5pt,linecolor=magenta](0.8,1.65)(0.5,1.85)(0.2,2.0)
  \rput[bl]{0}(-0.2,0.0){$\sigma$}
  \rput[bl]{0}(1.25,0){$\psi$}
  \endpspicture
  \; = (-1) \;
  \pspicture[0.5](2.0,2.0)
 
\pscoil[coilarm=1pt,coilwidth=5pt,coilaspect=0,linewidth=1.5pt,linecolor=blue]
  (1.2,0.0)(1.2,2.0)
    \psline[linewidth=1.5pt,linecolor=magenta](0.4,0.0)(0.4,2.0)
  \rput[bl]{0}(0,0.0){$\sigma$}
  \rput[bl]{0}(1.45,0){$\psi$}
  \endpspicture
\end{equation}
\begin{equation}
  \label{eq:braiding3}
  \pspicture[0.5](1.8,2.0)
  \psline[linewidth=1.5pt,linecolor=magenta] (1,0.0)(1.0,2.0)
  \pscurve[linewidth=1.5pt,linecolor=magenta,border=4pt]
  (0.2,0.0)(1,0.5)(1.4,1.1)(1.2,1.4)
  \pscurve[linewidth=1.5pt,linecolor=magenta](0.8,1.65)(0.5,1.85)(0.2,2.0)
  \rput[bl]{0}(-0.2,0.0){$\sigma$}
  \rput[bl]{0}(1.2,0){$\sigma$}
  \endpspicture
  \; = e^{-i{\pi}/{4}} \;
  \pspicture[0.5](2.0,2.0)
  \psline[linewidth=1.5pt,linecolor=magenta](1.2,0.0)(1.2,2.0)
  \psline[linewidth=1.5pt,linecolor=magenta](0.4,0.0)(0.4,2.0)
 
\pscoil[coilarm=1pt,coilwidth=5pt,coilaspect=0,linewidth=1.5pt,linecolor=blue]
  (0.4,1.0)(1.2,1.0)
  \rput[bl]{0}(0,0.0){$\sigma$}
  \rput[bl]{0}(1.4,0){$\sigma$}
  \rput[bl]{0}(0.7,0.45){$\psi$}
  \endpspicture
\end{equation}
\end{subequations}
We emphasize that these diagrams merely keep track of particle fusion
and braiding statistics. There are no additional propagators or
interactions associated with these diagrams that need to be
calculated, and these relations are unchanged by any smooth
deformations in which worldlines do not cross. The signature of
non-Abelian statistics is apparent in the third relation,
Eq.~(\ref{eq:braiding3}), where winding two $\sigma $ particles around
each other is seen to be equivalent (up to a phase) to
exchanging a $\psi$ particle between them.

Each quasihole in the MR state carries a $\text{U}(1)$ charge/flux of
$\left( {e}/{4},{{\Phi _{0}}}/{2}\right)$ as well as the Ising label
$\sigma$. A straightforward application of the fusion rules determines
that a composite of $n$ quasiholes will have $\text{U}(1)$ charge/flux
$\left(n {e}/{4}, n {{\Phi_{0}}}/{2}\right)$ and Ising label $Q_{n}$,
where $Q_{n}$ must equal $\sigma$ when $n$ is odd, but can equal
either $\mathbb{I}$ or $\psi$ when $n$ is even. We can combine the
braiding rules of the two sectors to get the rules for winding a
single quasihole counterclockwise around $n$ quasiholes by making the
following modifications to the diagram equations of the Ising sector
Eqs.~(\ref{eq:braiding}): assign $\left( {e}/{4},
  {{\Phi_{0}}}/{2}\right)$ to the leftmost and $\left(n {e}/{4},
  n{{\Phi _{0}}}/{2}\right) $ to the rightmost worldlines on each
side, assign $(0,0)$ to the $\psi$ worldline in
Eq.~(\ref{eq:braiding3}), and multiply the right hand side of each
equation by $\exp \left( i n {\pi }/{4}\right)$. These rules agree
with those obtained by explicitly manipulating quasihole wavefunctions
in the MR state \cite{Nayak96c}.

The inner product for the interference term $\left\langle \Psi \right|
M_{n}\left| \Psi \right\rangle$ is represented diagrammatically by the
standard closure, where each worldline is looped back onto itself in a
manner that introduces no additional braiding. From
Eq.~(\ref{eq:braiding3}), we find that if there is an odd number of
quasiholes on the antidot, $\left\langle \Psi \right| M_{n}\left| \Psi
\right\rangle$ is proportional to the following diagram (leaving
$\text{U}(1)$ labels implicit):
\begin{equation}
  \label{eq:link1}
  \pspicture[0.5](2.4,1.3)
  \psarc[linewidth=1.5pt,linecolor=magenta] (1.6,0.7){0.5}{180}{360}
  \psarc[linewidth=1.5pt,linecolor=magenta] (0.9,0.7){0.5}{0}{180}
  \psarc[linewidth=1.5pt,linecolor=magenta,border=4pt]
(0.9,0.7){0.5}{180}{360}
  \psarc[linewidth=1.5pt,linecolor=magenta,border=4pt] (1.6,0.7){0.5}{0}{180}
  \rput[bl]{0}(0.15,0.3){$\sigma$}
  \rput[bl]{0}(2.15,0.3){$\sigma$}
  \endpspicture
   \; = e^{i(n-1)\frac{\pi}{4}} \;
  \pspicture[0.5](3.3,1.3)
  \pscircle[linewidth=1.5pt,linecolor=magenta] (0.9,0.7){0.5}
  \rput[bl]{0}(0.15,0.3){$\sigma$}
  \pscircle[linewidth=1.5pt,linecolor=magenta] (2.6,0.7){0.5}
 
\pscoil[coilarm=1.2pt,coilwidth=5pt,coilaspect=0,linewidth=1.5pt,
linecolor=blue]
  (1.4,0.7)(2.1,0.7)
  \rput[bl]{0}(3.15,0.3){$\sigma$}
  \rput[bl]{0}(1.7,0.25){$\psi$}
  \endpspicture
\end{equation}
But this diagram has vanishing
amplitude as a result of the following general consistency condition in anyon
models:
\begin{equation}
  \label{eq:exclusion}
  \pspicture[0.5](2.0,2.0)
  \psline[linewidth=1.5pt,linecolor=black] (1,0.0)(1.0,0.5)
  \psline[linewidth=1.5pt,linecolor=black] (1,1.5)(1.0,2.0)
  \pscircle[linewidth=1.5pt,linecolor=black] (1,1.0){0.5}
  \rput[bl]{0}(1.2,0.0){$a$}
  \rput[bl]{0}(1.2,1.8){$b$}
  \rput[bl]{0}(0.2,0.9){$c$}
  \rput[bl]{0}(1.6,0.9){$d$}
  \endpspicture
   \; = \delta_{a,b} \;
  \pspicture[0.5](2.0,2.0)
  \psline[linewidth=1.5pt,linecolor=black] (1,0.0)(1.0,0.5)
  \psline[linewidth=1.5pt,linecolor=black] (1,1.5)(1.0,2.0)
  \pscircle[linewidth=1.5pt,linecolor=black] (1,1.0){0.5}
  \rput[bl]{0}(1.2,0.0){$a$}
  \rput[bl]{0}(1.2,1.8){$a$}
  \rput[bl]{0}(0.2,0.9){$c$}
  \rput[bl]{0}(1.6,0.9){$d$}
  \endpspicture
\end{equation}
where the labels indicate particle types permitted by the fusion
rules.  Thus, with no interference, we have
\begin{equation}
  \label{eq:interf-odd}
  \sigma _{xx}\varpropto \left| t_{1}\right| ^{2}+\left| t_{2}\right|^{2},
  \qquad  n \text{ odd}
\end{equation}

When there is an even number $n$ of quasiholes on the antidot, the
environment will effectively measure the state, forcing it into either
an overall $\mathbb{I}$ or $\psi$ (not a superposition of the two).
It is easy to see from Eqs.~(\ref{eq:braiding1},\ref{eq:braiding2}) that the
interference term $\left\langle \Psi \right| M_{n}\left| \Psi
\right\rangle = (-1)^{N_\psi}e^{in\pi/ 4}$, and thus
\begin{equation}
  \label{eq:interf-even}
  \sigma _{xx}\varpropto \left| t_{1}\right|^{2}
  +\left| t_{2}\right| ^{2} + (-1)^{N_\psi} 2\left|
    t_{1}\right| \left| t_{2}\right| 
  \cos \left( \beta +n\frac{\pi }{4}\right),
  \quad n \text{ even}
\end{equation}
where $\beta = \alpha + \text{arg}(t_2/ t_1)$, and so can be varied by
changing $B$ and the relative tunneling phase. Here, $N_\psi = 1$ when
the $n$ quasiholes are in the $\psi$ state and 0 otherwise.  We note,
that for two well-separated quasiholes, the energies of the two
possible combined states ($\mathbb{I}$ or $\psi$) are \emph{equal}.
This, however, is not going to be the case for two quasiholes on the
same antidot. In particular, one can write down the two corresponding
wavefunctions for the case of a ``small antidot'' with two
quasiholes are located at the origin \cite{Read96}:
  \begin{equation}
    \label{eq:MR2hI}
    \Psi_{2\text{qh},\mathbb{I}} 
    =  \prod_j z_j \; \Psi_{\text{GS}}
  \end{equation}
where
\begin{equation}
  \label{eq:MRGS}
  \Psi_{\text{GS}} =  \mathcal{A}\left(\frac{1}{z_1-z_2}
      \frac{1}{z_3 - z_4} \ldots\right)
\prod_{j<k} (z_j - z_k)^2 \prod_j e^{-|z_j|^2/4}
\end{equation}
is the
 ground state wave function for the MR state
with $\mathcal{A}(\ldots)$ denoting the antisymmetrized sum over all
possible pairings of electron coordinates, and
  \begin{multline}
    \label{eq:MR2he}
    \Psi_{2\text{qh},\psi} = \prod_j z_j \;
    \mathcal{A}\left(\frac{z_1-z_2}{z_1 z_2}
      \frac{1}{z_3 - z_4}\frac{1}{z_5 - z_6} \ldots\right)\\
    \times \prod_{j<k} (z_j - z_k)^2 \; \prod_j e^{-|z_j|^2/4}.
  \end{multline}
  While these wavefunctions are clearly different, using them as
  variational functions to estimate the energy difference for the case
  of realistic electron-electron interactions appears hopeless since
  even the ground state wavefunction (\ref{eq:MRGS}) is not actually a
  ground state for any such realistic interaction. While this remains
  an open problem, a very na\"{\i}ve argument would suggest that the
  energy difference should scale as $e^2/R$ where $R$ is the antidot
  radius; however, it is entirely possible that such a term will have
  a small prefactor. If charging the antidot is done adiabatically,
  one may hope that upon addition of two new quasiholes, the system
  will remain in the same energetically preferred state (probably
  $\mathbb{I}$). In such a case, $\sigma_{xx}$ is expected to cycle
  through all four possible values given by Eq.~(\ref{eq:interf-even})
  as a function of an increasing even number of quasiholes, while it
  returns to the same value given by Eq.~(\ref{eq:interf-odd}) for any
  odd number of quasiholes. However, if the combined state of an even
  number of quasiholes is chosen randomly every time, we cannot expect
  such even number periodicity, though the magnitude of the current
  will generically change whenever two quasiholes are added. The real
  test for the non-Abelian nature is done by changing the magnetic
  field $B$ at fixed filling fraction, for a various number of
  quasiholes on the antidot. In doing so, Aharonov-Bohm oscillations
  with period $4\Phi_0$ should be observed in the even $n$ case and no
  oscillations whatsoever should be seen for the odd $n$ case
  \footnote{While this Letter was in preparation, we learned about a
    similar work by Stern and Halperin \cite{Stern06a}.  Reaching the
    same conclusions, they offered an alternative method for detecting
    the quasihole parity, namely by changing the shape of the edge,
    effectively modifying the Abelian phase $\beta$ in our
    Eq.~(\ref{eq:interf-even}).}.

\paragraph{Implications for a topological qubit scheme}

\begin{figure}[hbt]
\includegraphics[width=2.7in]{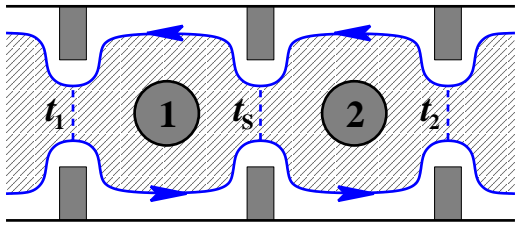}
\caption{The configuration for a topologically protected qubit
  proposed in \cite{DasSarma05}. A two-point interferometer is used to
  measure the combined state of a quasihole pair split onto two
  separate antidots.  A bit flip that switches between the
  $\mathbb{I}$ and $\psi$ states is performed by tunneling a single
  quasihole through the switching constriction, whose tunneling
  amplitude $t_{\text{S}}$ can be turned on and off by controlling the
  middle set of gates.}
\label{fig:qubit}
\end{figure}

We finally turn to the implications of our results to the proposed
implementation of a topological qubit \cite{DasSarma05}, which is
schematically shown in Fig.~\ref{fig:qubit}.

Assuming the qubit is implemented as prescribed, one nevertheless has
to address the issue that `stray' quasiparticles may disrupt the
ability to both measure and switch the state of the qubit. These stray
excitations may be trapped elsewhere in the system by a local disorder
potential. From Eqs.~(\ref{eq:braiding}) and the related discussion,
it is clear that in order to be able to detect the state of the qubit,
the \emph{total} number of quasiparticles (and quasiholes), including
the strays, in the area between the ``measurement'' tunneling contacts
($t_1$ and $t_2$ in Fig.~\ref{fig:qubit}) must be
\emph{even}, otherwise the interference necessary to distinguish the
states will not be seen.  Similarly, in order for switching to work,
the total number of quasiparticles in the left partition (i.e.
between $t_1$ and $t_{\text{S}}$ in Fig.~\ref{fig:qubit})
must be \emph{odd}, otherwise the state would simply acquire an
Abelian phase.

\paragraph{Conclusion} To summarize, in this letter we propose an
interferometric experiment for detecting non-Abelian quasiparticle
statistics in the MR state, the leading candidate for the $\nu=5/2$
FQHE plateau. Interestingly, while performing this experiment at
$\nu=5/2$ is expected to be more difficult than for well established
Laughlin states due to the smaller energy gap, the signature of a
non-Abelian state would be much easier to interpret due to the clear
separation of non-Abelian statistics from other effects that can only
contribute Abelian phases. The experimental setup
discussed here, while simpler than that recently proposed for a
topological qubit, may be a first step in the implementation of that
scheme.

\begin{acknowledgments}
  The authors would like to thank W.~Bishara, N.~Bonesteel, C.~Chamon,
  J.~Eisenstein, E.~Fradkin, F.~D.~M.~Haldane, S.~Simon, N.-C.~Yeh and
  especially C.~Nayak for many illuminating discussions. This work was
  supported in part by the NSF under Grant No.~EIA-0086038 and the ARO
  under Grant No.~W911NF-04-1-0236. K.~S.\ would also like to
  acknowledge the hospitality of the Aspen Center for Physics.
\end{acknowledgments}

\bibliography{../bibs/corr}

\end{document}